# The World Space Observatory Project

# WSO/UV http://wso.vilspa.esa.es/

I. Pagano[1], M. Rodonò[2], G. Bonanno[1], L. Buson[3], A. Cassatella[4], D. De Martino[5], W. Wamsteker[6], B. Shustov[7], M. Barstow[8], N. Brosch[9], Cheng Fu-Zhen[10], M. Dennefeld[11], A.I. Gomez de Castro[12], N. Kappelmann[13], J. Sahade[14], K. Van der Hucht[15], J.-E. Solheim[16], H. Haubold[17], A. Altamore[19], V. Andretta[5], M. Badiali[4], U. Becciani[1], I. Busà[1], E. Cappellaro[5], D. Cardini[4], S. Catalano[1], V. Castellani[27], ,M. Chiaberge[25], A. Chieffi[4], C. Chiuderi[20], R. Cosentino[1], G. Cremonese[3], G. Cutispoto[1], R. Falomo[3], F. Ferrini[27], M.G. Franchini[23], A. Frasca[1], F. Giovannelli[4], L. Gori[20], M.T. Gomez[5], M. Hack[24], A.F. Lanza[1], A. Lanzafame[2], M.L. Malagnini[24], E. Marilli[1], P. Marziani[3], F. Matteucci[24], C. Morossi[23], U. Munari[3], E. Pace[20], N. Panagia[26], L. Pasinetti[21], G. Piotto[3], F. Polcaro[4], M. Radovich[5], S. Ragaini[5], A. Rifatto[5], C. Rossi[18], S. Scuderi[1], P. Selvelli[4], R. Silvotti[5], L. Terranegra[5], M. Turatto[3], M. Uslenghi[22], R. Viotti[4]

1) INAF-Catania, **Italy**; 2) Univ. Catania, **Italy**; 3) INAF-Padova, **Italy**; 4) CNR/IASF-Roma, **Italy**; 5) INAF-NA, **Italy**; 6) **ESA**-Vilspa; 7) INASAN, **Russia**; 8) Univ. Leicester, **UK**; 9) Wise Obs., Israel; 10) USTC-CfA, **China** 11) IAP, **France**; 12) UCM, **Spain**; 13) Univ. Tübingen, **Germany**; 14) CONAE, **Argentina**; 15) Sron-HEA, **Netherlands**; 16) Tromsø Univ., **Norway**; 17) **UN**-OOSA; 18) Univ. Roma 1, **Italy**; 19) Univ. Roma 3, **Italy**; 20) Univ. Florence, **Italy**; 21) Univ. Milano, **Italy** 22) CNR/IASF-Milano, **Italy**; 23) INAF-Trieste, **Italy**; 24) Univ. Trieste, **Italy**; 25) CNR/IRA-Bologna, **Italy**; 26) ESA-STScI, **USA**; 27) Univ. Pisa, **Italy**.

## Background

❖ The World Space Observatory (WSO) concept was discussed for the first time in the conclusions and recommendations of the 8th UN/ESA Workshop for Basic Space Science in the Developing Countries.

❖ An ESA internal study was proposed in order to assess the mission feasibility and to provide the conceptual design of the WSO/UV space/ground system. The study was financed under the ESA General Studies Programme, under the responsibility of W. Wamsteker (ESA):

❖ **ESA-CDF assessment study ESA/CDF-05(A)**
*http://wso.vilspa.esa.es/docs/WCC/DOC/Attachments/GEN-TN-0002-1-0.pdf (May 2000)*

This was followed by

❖ A **JPL/NASA assessment study ADP Report: CL#01-1168**
*http://wso.vilspa.esa.es/docs/WCC/DOC/Attachments/GENTN-0001-Draft-0.pdf (Gen 2001)*

## WSO/UV facts

**WSO/UV is an International Collaboration to build a UV (103-310 nm) dedicated telescope (1.7m) capable of:**

- ✓ high resolution spectroscopy
- ✓ long slit low resolution spectroscopy
- ✓ deep UV imaging
- ✓ free of visibility constraints (L2)
- ✓ "real time" operations
- ✓ investigate all possible time-scales

## WSO/UV principia

❖ Use application innovation, but avoid technical innovation
❖ Use heritage as much as possible;
❖ Apply new engineering methods (*concurrent design*);
❖ Keep the mission simple;
❖ Science Operations Centers at National level.

## Performances

✓**Spectral resolution**: WSO/UV-HIRDES ≈ HST-STIS

WSO/UV-HIRDES > HST-COS

✓**Sensitivity**: WSO/UV-HIRDES ≈ 5-10 × HST-STIS

WSO/UV-HIRDES ≈ HST-COS

✓WSO/UV is a **dedicated UV** telescope

✓WSO/UV has a **high efficiency** of observations **at L2**

**WSO/UV will provide a net increase in UV productivity of a factor ~40-50 compared to HST/STIS.**

## WSO/UV-HIRDES vs. HST-STIS

**High Resolution**
λ/Δλ 50,000 WSO/UV (3 pix.)
λ/Δλ 37,000 HST/STIS (2 pix.)

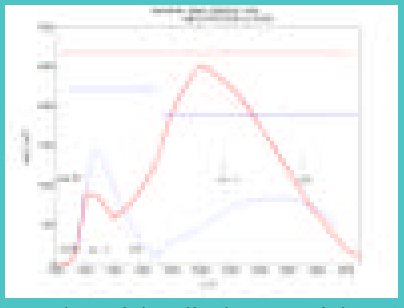

**Low Resolution**
λ/Δλ 1,000 WSO/UV (3 pix.)
λ/Δλ 1,000 HST/STIS (2 pix.)

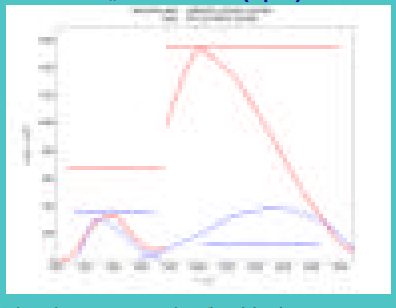

Comparison of the Effective Area of the WSO/UV instruments (*red*) with the UV spectrographs STIS (*blue*) on HST at comparable resolution. The horizontal dashed lines give the Free Spectral Range (i.e. the I range which can be observed in a single exposure) of the spectrographs. The wavelengths of the Lyman a and CIV lines at redshift of 0.75 are indicated in the right hand figure.

## Science

WSO/UV will allow us to:

✓ observe objects 4-5 magnitudes fainter than possible with HST, providing completely new opportunities in **extragalactic astronomy and cosmology**.

✓ Carry out large scale, **high resolution spectroscopic surveys** of galactic sources.

With a 2007/8 launch date, WSO is ideally placed to provide follow-up studies of the large number of UV sources expected from the GALEX sky survey.

## Main WSO/UV Science Interests of the Italian UV community

- ❖ **Hot stars and mass loss phenomena, Supernovae (STScI, CT)**
- ❖ **Hot horizontal branch stars (PI)**
- ❖ **Interacting Binaries (PD, RM)**
- ❖ **Novae (TS, PI)**
- ❖ **Cool Stars, Atmospheric Structure, Magnetic Activity (CT, TS)**
- ❖ **UV-bright stars, Young Stars in Globular Clusters (PI, RM, CT)**
- ❖ **Stellar populations in Galaxies (NA)**
- ❖ **Active Galactic Nuclei (MI)**
- ❖ **Cataclysmic Variables (NA, RM)**
- ❖ **Low Mass X-ray Binaries (NA, PI)**
- ❖ **Interstellar medium (PI)**

## Telescope and Payloads

Lavochkin Association

### The telescope is heritage of the Spectrum-UV Project.

**To be used by WSO/UV it was necessary to have optical modifications, mass reduction, cost reduction.**

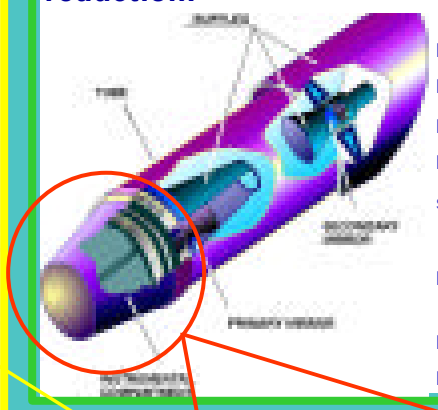

| | |
|---|---|
| D | 170 cm |
| F | 1700 cm |
| F/10 | |
| FoV | 30 arcmin (150 mm) |
| scale | 12.05 arcsec/mm |
| Imaging | |
| $D_{80}$ ® | 0.35 arcsec @ 633 nm |
| $D_{80}$ ® | 0.07 arcsec @ 122 nm |

### The Imaging Camera design is based on the Tauvex mission.

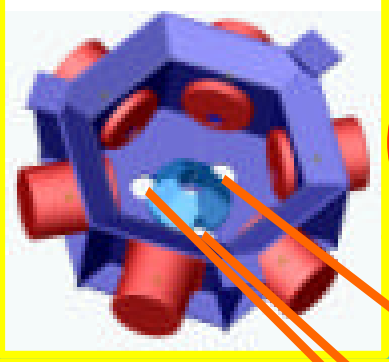

A Central Field Camera will sample the diffraction limited resolution. Additional imagers can be allocated by using 90° single reflections:

❖ 2 high resolution UV cameras

❖ 2 high sensitivity UV cameras

❖ 2 optical imagers.

A study for a polarimeter unit is in progress.

Assuming that:
a) the telescope has 0.07" (80% Encircled Energy);
b) the MCP diameter is 25 mm;
c) the read-out anode has 2000x2000 pixel => pixel size = 12.5 micron

| Camera | Aperture | velength Ran | SF Sampling | FOV | Photocatode | Number of mirrors |
|---|---|---|---|---|---|---|
| Central Field | F/10 | 120-190 nm | 0.15"/pixel | 5 arcmin | CsI | 0 |
| UV Imager HS 1 | F/10 | 120-290 nm | 0.15"/pixel | 5 arcmin | CsTe | 1 |
| UV Imager HR 1 | F/50 | 120-290 nm | 0.03"/pixel | 1 arcmin | CsTe | 2 |
| UV Imager HS 2 | F/10 | 120-290 nm | 0.15"/pixel | 5 arcmin | CsTe | 1 |
| UV Imager HR 2 | F/50 | 120-290 nm | 0.03"/pixel | 1 arcmin | CsTe | 2 |
| Optical Imager HS | F/10 | 350-550 nm | 0.15"/pixel | 5 arcmin | Bialkali | 1 |
| Optical Imager HR | F/50 | 350-550 nm | 0.03"/pixel | 1 arcmin | Bialkali | 2 |

The reflection of Al + $MgF_2$ is about 0,85 120 - 290 nm; 0,9 350 - 500 nm

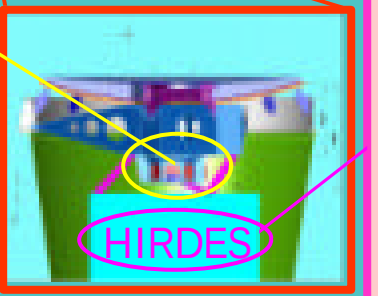

❖ Each of the three spectrometers has its own entrance slit

❖ The three optical trains are used in sequential mode

❖ This is managed by satellite motion with a pointing stability requirement of 0.1 arcsec to be monitored by three **Fine Guidance Sensors**.

### HIRDES: the High Resolution Double Echelle Spectrograph is heritage of the German Orpheus missions.

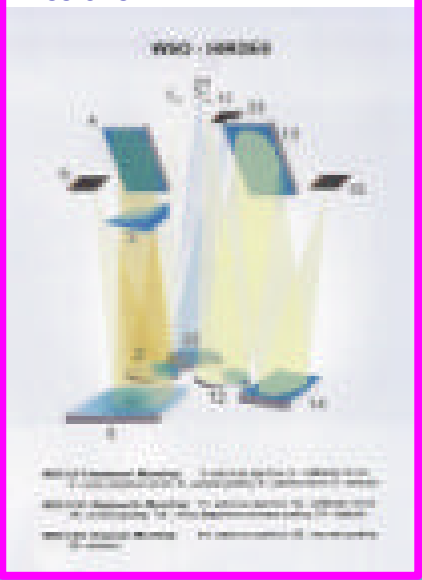

**Technical details of the UV Echelle Spectrograph UVES – 174.5-310.0 nm**

**Entrance aperture** circular 80 μm
**Collimator mirror** toroidal, R1 = 1608 mm, R2 = 1593 mm, circular 80 mm
**Echelle grating** 40 grooves/mm, 66.9° blaze, 90 mm x 190 mm
**Cross dispersion prism** 12° quartz, double pass, 100 mm x 110 mm
**Camera mirror** spherical, R=1600 mm, 110 mm x 130 mm
**Detector**: 30 mm (echelle dispersion), 30 μm resolution (pixel-width)
40 mm (prism dispersion), 40 μm resolution (pixel-height)
**Resolving power** R = 50000 (3 pixel criterion)

**Technical details of the VUV Echelle Spectrograph - VUVES – 102.8-172.6 nm**

**Entrance aperture** circular 80 μm
**Collimator mirror** parabolic, R = 1600 mm, 6° -off-axis, circular 80 mm
**Echelle grating** 65 grooves/mm, 71° blaze, 90 mm x 270 mm
**Wadsworth grating** toroidal, 625 grooves/mm, R1 = 1698 mm, R2 = 1695 mm, 90 mm x 120 mm
**Detector**: 30 mm (echelle dispersion), 30 μm resolution (pixel-width)
40 mm (Wadsworth dispersion), 40 μm resolution (pixel-height)
**Resolving power** R = 55000 (3 pixel criterion)

**Technical details of the Long Slit Spectrograph LSS – 102.8-310.0 nm**

**Entrance aperture** rectangular, 80 μm x 6 mm
**Rowland grating** aberration corrected concave grating, R 800 mm
**Detector**: 80 mm (grating dispersion), 15 μm resolution (pixel-width)
**Resolving power** examples: R = 1000 (3 pixel criterion), 110 nm to 350 nm

### Fine Guidance System (FGS)

• In the WSO/UV design, the FGS units are part of the focal plane instruments.

• The FGS are required to maintain during science observation a pointing error < 0.03 arcsec (1 σ) over a period of 24 hours on all three axes

• To be able to support the tracking of objects moving up to 0.2 arcsec per second

**FGS Requirements:**

- Wavelength Range: 400nm .. 800nm (visible)
- Detector: 1024 x 1024 Pixel x16bit, CCD, pixel size 24um squared, overall active size 25mm x 25mm
- FOV: 55.2 arcsec square field extension
- IFOV = 0.054 arcsec
- 3 FGS arranged symmetrically around the optical axis with a distance of 20mm of each detector to the optical axis

## WSO/UV Organizational status

The activities are coordinated by a WSO/UV Implementation Committee (WIC)

**Non European Members: Russia (chair), Argentina, Cina, Israele, Messico, Sud Africa, Ucraina**

**European Members: Francia, Germania, Italia, Olanda, Paesi Nordici (Danimarca, Finlandia, Lituania, Norvegia, Svezia), Spagna, Regno Unito**

**Others: ESA, UN**

**14 NWWG's** representing scientists form 17 countries.

The total membership of the NWWG's comprise **130** scientists from ESA member states; **70** scientists in non-member states; and some **10** industries.

## Possible Italian Contribution to Phase A Study proposed to ASI

➢ **Detectors** (*Photon Counter Intensified Active Pixel Sensors PC-IAPS*)

Catania (INAF), Milano (CNR), Laben (elettronica)

➢ **Fine Guidance System**

Galileo Avionica, Dip. Astronomia di Firenze XUVLab, Laben

➢ **Instruments Control & Data Processing Units (ICU/DPU)**

Laben

➢ **Scientific Objectives Definition**

Researchers at several Institutions

1. Selection of filters for UV and optical cameras
2. Definition of strategies for pre/in flight calibration
3. Participation in the edition of the book "Science with WSO/UV" (ed. M. Barstow – Leicester Univ., UK)

## BUS

**The bus developed for Herschel/Planck (and Eddington), by Alenia Spazio, has been selected by ESA as a working hypothesis for WSO/UV.**

## LAUNCHER

❖ **Middle class launcher Soyuz-2, or Zenith2E with booster Fregat are capable to put into L2 point vicinity the S/C up to 3.6 ton**

❖ **Chinese Long-March launcher is under consideration.**

## L2 Lagragian orbit

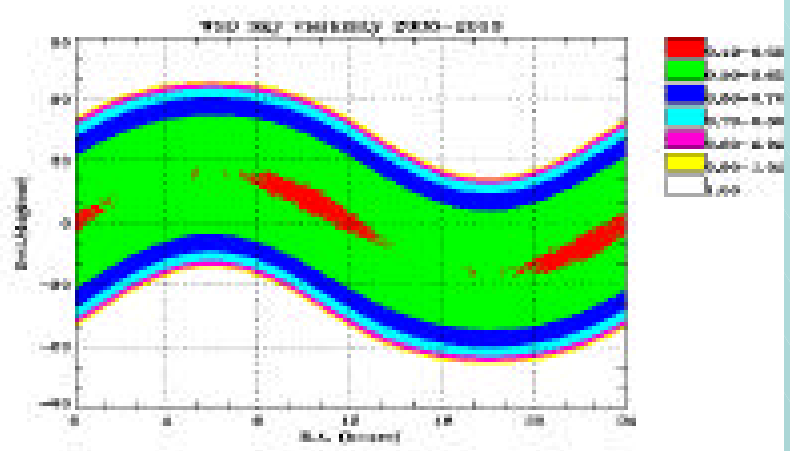

## Implementation Plan

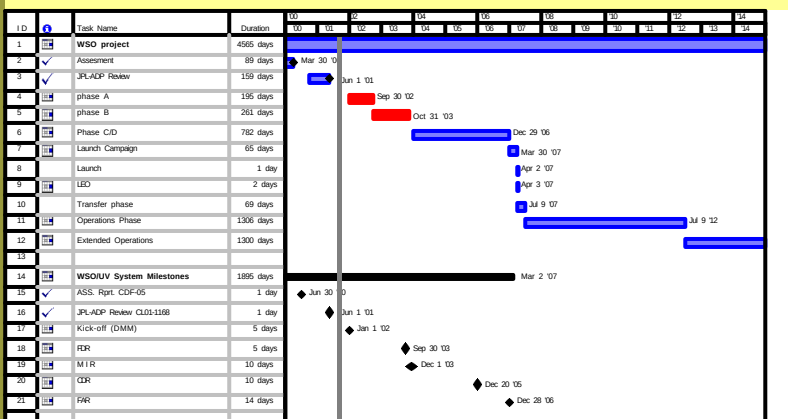